\documentclass[letter]{aa}     

\usepackage{graphicx}
\usepackage{txfonts}
\usepackage{amsmath}
\usepackage[colorlinks,allcolors=blue,bookmarks=false,hypertexnames=true]{hyperref} 
\usepackage{xcolor}
\usepackage{placeins}

\newcommand{\HI}{H\,{\sc i} }
\newcommand{\msol}{\mbox{$ \rm M_\odot$} }

\newcommand {\kms}{\ifmmode{\rm km \, s^{-1}}\else{$\rm km \, s^{-1}$}\fi}

\realpagewiselinenumbers
\begin{document}

   \title{Discovery of an isolated RELHIC Candidate: J1351+0039}

   \subtitle{}

%
%
%

   \author{Jin-Long Xu\inst{1,2}\corrauth{xujl@bao.ac.cn}        
        \and Alejandro Ben{\'\i}tez-Llambay\inst{3}\corrauth{alejandro.benitezllambay@unimib.it}
        \and Ming Zhu\inst{1,2}\email{mz@nao.cas.cn}
        \and Chuan-Peng Zhang\inst{1,2}\email{cpzhang@nao.cas.cn}
        \and Nai-Ping Yu\inst{1,2}\email{npyu@bao.ac.cn}
        \and Xiao-Lan Liu\inst{1,2}\email{liuxiaolan@bao.ac.cn}
        \and Mei Ai\inst{1,2}\email{aimei@nao.cas.cn}
        \and Ruilei Zhou\inst{1,2,4}\email{zhourl@bao.ac.cn}
        \and Bin Liu\inst{1,2}\email{bliu@nao.cas.cn}
        }

   \institute{ State Key Laboratory of Radio Astronomy and Technology, National Astronomical Observatories, Chinese Academy of Sciences, Beijing 100101, China
   \and Guizhou Radio Astronomical Observatory, Guizhou University, Guiyang 550000, China
   \and Dipartimento di Fisica G. Occhialini, Universit\`a degli Studi di Milano Bicocca, Piazza della Scienza, 3 I-20126 Milano MI, Italy
   \and University of Chinese Academy of Sciences, Beijing 100049, China}

   \date{Received xx xx, 20XX}

 
  \abstract
   {We report the discovery of J1351+0039, a new isolated dark galaxy candidate identified through a blind \HI survey with the Five-hundred-meter Aperture Spherical radio Telescope (FAST), combined with deep optical imaging from the DESI Legacy Survey. The object displays a compact \HI structure with a systemic velocity of \( V_{\mathrm{LSR}} = 429.0 \pm 1.4 \, \mathrm{km\,s^{-1}} \) and an inferred \HI mass of \( (4.5 \pm 1.7) \times 10^{5} \, \rm M_{\odot} \) at a distance of 3.6 $\pm$ 0.7 Mpc.  No optical counterpart is detected in the DESI  images, with a $g$-band surface brightness limit of 28.1  mag arcsec$^{-2}$, corresponding to a stellar mass limit of \( M_{\star} \lesssim 1.2 \times 10^{5} \, \rm M_{\odot} \). The dynamical mass is estimated to be \( M_{\mathrm{dyn}} \approx 1.3 \times 10^{8} \, M_{\odot} \), yielding a dark-to-baryonic mass ratio of \( \sim 200 \), implying that J1351+0039 is overwhelmingly dominated by dark matter. While   weak rotation is detected in J1351+0039, the kinematics are  dominated by internal velocity dispersion, confirming pressure support and reminiscent of Reionization-Limited \HI Clouds (RELHICs). J1351+0039 stands out as one of the most promising isolated RELHIC candidates, offering a unique laboratory for studying baryonic physics and dark matter in the absence of star formation.}

   \keywords{galaxies: dwarf – galaxies: formation – galaxies: kinematics and dynamics – dark matter
               }

   \maketitle

\section{Introduction}

In the  Lambda Cold Dark Matter ($\Lambda$CDM) hierarchical structure formation scenario, dark matter halos assemble first, after which gas accretes into these gravitational potentials and may eventually cool and form stars. However, not all dark matter halos are expected to host a luminous  galaxy.  In particular, low-mass halos may retain their primordial gas without ever forming stars, owing to inefficient cooling prior to reionization, or to photoheating by the cosmic ultraviolet background (UVB) after reionization \citep[e.g.,][and references therein]{2017ARA&A..55..343B,2020MNRAS.498.4887B}. In the $\Lambda$CDM paradigm, such gas-bearing, starless dark matter halos are usually referred to as Reionization-Limited \HI Clouds \citep[RELHICs;][]{2017MNRAS.465.3913B,  2024ApJ...962..129L, 2026A&A...710A.156G}, or simply ``dark galaxies’’. Their existence provides a critical  test of hierarchical structure formation models and the physics of galaxy formation at the smallest scales \citep[see, e.g.,][]{2022NatAs...6..897S}.

Despite being a robust theoretical prediction, the observational identification of dark galaxies has proven remarkably challenging. Because they lack a detectable stellar population, dark galaxies cannot be identified through conventional optical or near-infrared surveys. Instead, they must be discovered via the 21-cm emission line of neutral hydrogen, which serves as the primary tracer of their baryonic content. As catalogs of \HI sources grow \citep{2025ApJS..279...38K,2026MNRAS.548ag732M,2026arXiv260714584X}, it is crucial to distinguish between observationally faint dwarf galaxies, which appear dark solely due to surface-brightness limits, and genuinely starless halos (RELHICs). Cosmological simulations demonstrate that RELHICs are low-mass, quasi-spherical, and largely pressure-supported systems where the gas remains in hydrostatic equilibrium with the halo potential and in thermal equilibrium with the cosmic ultraviolet background \citep[e.g.,][]{2017MNRAS.465.3913B,  2024ApJ...962..129L, 2026A&A...710A.156G}. The discovery of Cloud-9 near the spiral galaxy M94 provided the first possible observational confirmation of this pristine, starless class \citep{2023ApJ...952..130Z,2023ApJ...956....1B,2025ApJ...993L..55A,2026MNRAS.550g1293Z}, followed by two additional RELHIC candidates near M51 \citep{2026A&A...712L..12C}.

Although several RELHIC candidates have been identified  to date, there remains a lack of confirmed isolated RELHICs supported by detailed follow-up studies. Isolated candidates are particularly valuable because they can be directly ruled out as tidal debris, and their isolation minimize uncertainties of the influence of the the background medium on estimates of dark matter content \citep[see, e.g.,][]{2026A&A...712A..11T}. In this Letter, we report the discovery of a new isolated dark galaxy candidate identified through a blind \HI survey conducted with the Five-hundred-meter Aperture Spherical radio Telescope (FAST) \citep{2024SCPMA..6719511Z,2026SCPMA..6929811Z}, combined with deep optical imaging from the Dark Energy Spectroscopic Instrument (DESI).

\begin{table}
\caption{Measured and derived  properties of J1351+0039. We list: equatorial coordinates (R.A., Decl.); line widths  at 50\% of the peak flux ($W_{50}$); line widths  at 20\% of the peak flux ($W_{20}$);  system velocity ($V_{\rm LSR}$); total flux ($S_{v}$); distance ($D$);  effective radius ($R_{\rm HI}$);  limiting stellar mass ($M_{\rm \star}$); \HI gas mass ($M_{\rm HI}$);  dynamic mass ($M_{\rm dyn}$).}
\centering
\begin{tabular}{lcc}
\hline\hline
R.A. & 13$^{\rm h}$51$^{\rm m}$33.4$^{\rm s}$  \\
Decl. & 00$^{\rm \circ}$39$^{\rm \prime}$36.0$^{\rm \prime\prime}$ \\
$W_{\rm 50}$ ($\kms$) & 27.6$\pm$2.0  \\
$W_{\rm 20}$ ($\kms$) & 32.4$\pm$3.2   \\
$V_{\rm LSR}$ ($\kms$) & 429.0$\pm$1.4 \\
$S_{v}$      (mJy km s$^{-1}$) & 147$\pm$10 \\ 
$D$ (Mpc)                     &  3.6$\pm$0.7\\
$R_{\rm HI}$ (kpc) & 2.8$\pm$0.5  \\
$M_{\rm \star}$ (\msol) & $<$1.2$\times10^{5}$ \\
$M_{\rm HI}$ (\msol) & (4.5$\pm$1.7)$\times10^{5}$  \\
$M_{\rm dyn}$ (\msol) & (1.3$\pm$0.3)$\times10^{8}$  \\
\hline
\end{tabular}
\end{table}

\section{Observations and data reduction}
Studies of dark and faint galaxies in the 21-cm \HI line (1420.4058 MHz) require  both high detection sensitivity  and high velocity resolution. Previously, using the FASHI Data Release 2 \citep[DR2;][]{2026SCPMA..6929811Z} , we found an isolated \HI cloud  with a detection sensitivity of 1.6 mJy/beam at a velocity resolution of 3.2 \kms.  To improve sensitivity and velocity resolution, we used FAST to observe the cloud three times during July-August 2026 \citep{Jiang+19,Jiang+20}. Mapping observations  were performed using the Multi-beam on-the-fly (OTF) mode. This mode  maps the sky with 19 beams simultaneously along consistent scanning  trajectories. The scan  rate  was set to 15$^{\prime\prime}$ s$^{-1}$, with an integration time of 1 second per spectrum.  We utilized the digital Spec(W) backend, which has a bandwidth of 500 MHz and 64k channels, yielding a frequency resolution of 7.629 kHz. The backend nominally operates over the frequency range  1050 MHz--1450 MHz.  To calibrate intensity to antenna temperature ($T_{\rm A}$), a noise signal with an amplitude of 10 K was injected over a period of 64 seconds. The pointing accuracy of the telescope was better than 10$^{\prime\prime}$.  The half-power beam width for each beam is $\sim$2.9$^{\prime}$ at 1.4 GHz. The bright quasar 3C 286 was used for absolute flux density ($S_{v}$)  calibration and  antenna gain correction. A gain  factor $T_{\rm A}/\it S_{v}$ was measured to be  approximately 13-17 K Jy$^{-1}$. The data reduction was performed by the Python-based pipeline HIFAST for FAST \citep{Jing+2024}. The calibrated standard  data cube has a pixel size of $1.0^{\prime}\times1.0^{\prime}$.  To construct a high-sensitivity \HI image, the velocity resolution of the FAST data was smoothed to 3.2 $\kms$. The mean  rms noise across the combined FASHI DR2 and new follow-up observations is $\sim$0.6 mJy beam$^{-1}$.

\section{Results}

\begin{figure}
        \centering
        \includegraphics[width=0.35\textwidth]{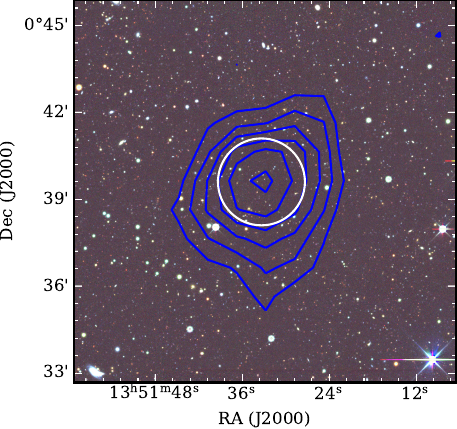}
        \includegraphics[width=0.35\textwidth]{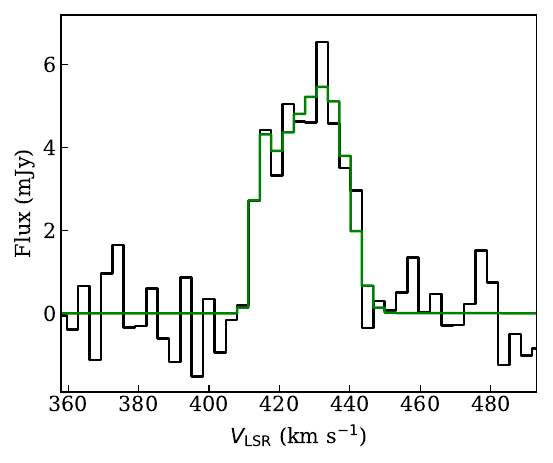}
        \vspace{-2mm}
        \caption{Upper panel: \HI column-density map of J1351+0039 from the FAST observation shown in blue contours, which is overlaid on the DESI RGB (g,r,z) image. The blue \HI contours begin at 6.3$\times$10$^{17}$ cm$^{-2}$ (3$\sigma$) in steps of 5.0$\times$10$^{17}$ cm$^{-2}$. The size of the white circle is 3$^{\prime}$, which coincides with the FAST beam. Lower panel: global \HI profile of J1351+0039 shown in a black line. The superposition of all signals with a column density greater than 3$\sigma$  and within the shown spatial extent. The green line indicates the BusyFit fitting result.}
        \label{FigGam1}%
\end{figure}

\begin{figure*}
        \centering
        \includegraphics[width=0.34\textwidth]{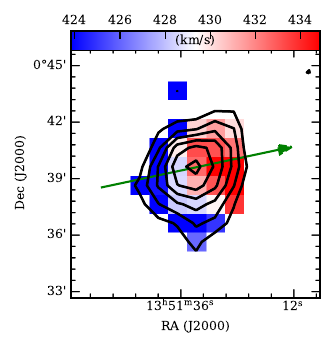}
        \includegraphics[width=0.34\textwidth]{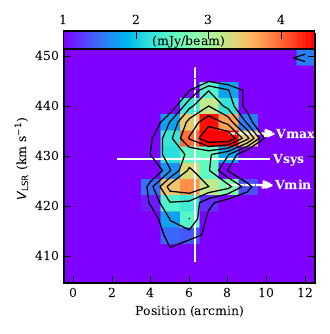}
        \vspace{-4mm}
        \caption{Left panel: velocity-field map of J1351+0039 in color scale overlaid with its \HI column-density map in black contours. The black contours begin at  6.3$\times$10$^{17}$ cm$^{-2}$ in steps of 5.0$\times$10$^{17}$ cm$^{-2}$.  Right panel: position-velocity (PV) diagrams of the observed data in color scale overlaid with the black contours. The black contours begin at 2$\sigma$ (1.2 mJy beam$^{-1}$) in steps of 1$\sigma$.  The cutting direction of the PV diagram is indicated by the blue arrow in left panel.  }
        \label{FigGam2}%
\end{figure*}

Based on the FASHI DR2, we identified an \HI cloud with a system velocity ($V_{\rm LSR}$) of 429.0 \kms. Its sky coordinates and velocity were derived from the \HI data cube using the Source Finding Application  \citep[SoFiA2,][]{2021MNRAS.506.3962W}.  Using the numerical action method (NAM) calculator \citep{2017ApJ...850..207S}, the inferred galactic standard of rest velocity ($V_{\rm gsr}$=377 \kms) and supergalactic coordinates (SGL=119.6983$^\circ$ and SGB=13.7797$^\circ$) places this cloud at a distance of $D = 3.6 \pm 0.7\text{ Mpc}$, which is  well associated with that (3.7 Mpc) from the Cosmicflows-4 distance calculator \citep{2020ApJ...902..145K}. The distance estimate for the NAM method has  an accuracy of 20\% within the Local Volume \citep{2024Ap.....67..433K}.  \cite{2013AJ....145..101K} presented an all-sky catalog of 869 nearby galaxies within 11 Mpc. Using this catalog, no galaxies lie within a projected angular separation of <3.0$^{\circ}$ (200 kpc at its distance) and system velocity of less than 250 \kms, implying that this cloud is an isolated object. Repeated observations at different epochs allowed us to rule out the possibility that the signal originates from radio-frequency interference (RFI) and also to confirm the object’s coordinates. In addition, we cross-checked with two candidate samples of dark galaxies \citep{2025ApJS..279...38K,2026MNRAS.548ag732M}, and found that our cloud is not in these tables. These results  establish the \HI gas cloud as a newly identified isolated source. Based on its coordinates, we designate it J1351+0039.

Figure \ref{FigGam1} (upper panel) presents the \HI column-density map of J1351+0039, shown as blue contours. The blue contours begin at \(6.3 \times 10^{17} \, \mathrm{cm^{-2}}\), corresponding to the \(3\sigma\) noise level. We constructed the column-density distribution map from the zeroth-moment (integrated intensity) map  over the velocity range 416–443 \kms. The \HI emission in the upper panel of Figure \ref{FigGam1}  exhibits a compact morphology, suggesting the presence of a gravitational center that has facilitated gas condensation. The effective radius of the compact gas  is determined as $R_{\rm eff} = \sqrt{D_{\rm maj}\times D_{\rm min}-B_{\rm size}^{2}}/2$, where \(D_{\rm maj}\) and \(D_{\rm min}\) are the major and minor diameters measured from the SoFiA-fitted ellipse to the \HI distribution above the \(3\sigma\) column-density threshold, which encompases about 95\% fraction of the total flux, and \(B_{\rm size}\) is the FAST beam size (\(2.9^{\prime}\)). The fitted \(D_{\rm maj}\) and \(D_{\rm min}\) are $\sim$6.4$^{\prime}$ and  $\sim$6.0$^{\prime}$, respectively, yielding an aspect ratio of 0.94. For J1351+0039, we derive an \HI radius of $\sim$2.8$^{\prime}$.

Figure \ref{FigGam1} (lower panel) displays the global \HI profile of J1351+0039, with the observed spectrum shown as a black line. The panel presents the integrated \HI spectrum obtained by summing all emission with column densities above the \(3\sigma\) threshold. We fitted the profile with a BusyFit function to derive key parameters \citep[see, e.g.,][]{2014MNRAS.438.1176W}. The fitted parameters are summarized in Table 1. The systemic velocity ($V_{\rm LSR}$) of J1351+0039 is \(429.0\pm1.4\) \kms. The total flux is \(147\pm10\) mJy km s\(^{-1}\), and the line widths at 50\%  and 20\%  of the peak flux are \(W_{50} = 27.6\pm2.0\) \kms and \(W_{20} = 32.4\pm3.2\) \kms. 

\begin{figure}
        \centering
        \includegraphics[width=0.35\textwidth]{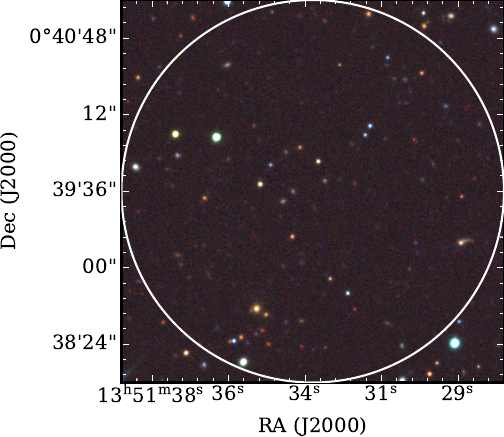}
        \vspace{-1mm}
        \caption{A smaller cutout of the image shown on the upper panel of Figure 1. The diameter of the white circle is 3$^{\prime}$. While some resolved galaxies or point-like sources are visible within the circle, no compact stellar overdensity is readily observed, and none of the galaxies coincide in redshift with J1351+0039.} 
        \label{FigGam3}%
\end{figure}

Figure \ref{FigGam2}  presents the velocity field and position–velocity (PV) diagram of J1351+0039. The left panel shows the velocity field in color, overlaid with \HI column-density isocontours (black lines) at the same levels as in Figure 1 (upper panel). The velocity field (Figure 2, left) reveals a mild, coherent velocity gradient of 4.5 \kms  across the $\sim 6'$ extent from southeast to northwest. The right panel of Figure 2 displays the PV diagram extracted along the major kinematic gradient axis (indicated by the green arrow). The PV slice shows a continuous velocity structure spanning from $V_{\rm min} \approx 424.8\text{ km s}^{-1}$ to $V_{\rm max} \approx 433.3\text{ km s}^{-1}$ centered on the systemic velocity ($V_{\rm sys} = 429.0\text{ km s}^{-1}$). The velocity gradient and velocity structure may be interpreted as rotation. We measure an observed  rotation velocity of \(V_{\rm rot} \simeq 4.2\) \kms, derived as \((V_{\rm max} - V_{\rm min})/2\), indicating that if there is rotation, it is also a weak rotation.  Likewise, we estimate the dispersion velocity as \(\sigma_v = \sqrt{W_{\rm 50}^{2} - V_{\rm rot}^{2}}/\sqrt{8ln(2)}\), yielding \(\sigma_v \simeq 11.6\) \kms.

To investigate the presence of  a stellar counterpart associated with J1351+0039, we used \(g\)-, \(r\)-, and \(z\)-band images from the DESI Legacy Survey (LS)\footnote{The DESI data are publicly available at the DESI Legacy archive:https://www.legacysurvey.org/}. Figure \ref{FigGam3} shows a smaller cutout image of the DESI LS, centered at the peak of the \HI emission. No optical extended emission is detected within this field, suggesting that any stellar counterpart, if present, is below the detection threshold of DESI LS. From the Figure \ref{FigGam1}, we see the \HI emission is larger than 3$^{\prime}$. In addition, we carefully inspected the entire \HI plane ($\sim6^{\prime}$) of J1351+0039  and found no associated extended optical emission.

\section{Discussion and Conclusion}
We have detected an isolated \HI gas cloud, J1351+0039, with a systemic velocity of $V_{\rm LSR} = 429.0 \pm 1.4\text{ km s}^{-1}$ and a compact, quasi-spherical morphology (\(D_{\rm min}/D_{\rm maj}\) = 0.94). Located at high Galactic latitude ($b \approx +59.8^\circ$), the field has negligible galactic extinction. The high systemic velocity and symmetric structure rule out a Galactic high-velocity cloud \citep{1997ARA&A..35..217W}. The inferred distance of 3.6 Mpc  places J1351+0039 well beyond the Local Group turnaround radius and far in the foreground of the Virgo Cluster ($D \approx 16.5\text{ Mpc}$). The absence of cometary tails, bridges, or nearby companions rules out a tidal origin.

At $D = 3.6\text{ Mpc}$, the \HI mass of J1351+0039  is $M_{\rm HI} = (4.5 \pm 1.7) \times 10^5 \rm ~M_\odot$, corresponding to a total gas mass of $M_{\rm gas} \approx 6.0 \times 10^5 \rm ~M_\odot$ including helium. Assuming no stars, we obtained  the baryonic mass, $M_{\rm bar} \approx 6.0 \times 10^5 \rm ~M_\odot$. Across the effective radius, $R_{\rm HI} = 2.8 \pm 0.5\text{ kpc}$, the mean gas surface density is very low ($\Sigma_{\rm gas} \approx 0.024 ~M_\odot\text{ pc}^{-2}$). In the absence of dark matter, balancing the observed velocity dispersion ($\sigma_v \approx 11.6\text{ km s}^{-1}$) against a self-gravitating gas disk would require an unphysical  vertical scale height exceeding the observed extent of the system by two orders of magnitude. Thus, without external gravitational confinement, the cloud would disperse on a sound-crossing timescale of $\sim 2.4 \times 10^8\text{ yr}$. Maintaining hydrostatic equilibrium therefore requires an external dark matter halo with an enclosed dynamical mass, which is computed using \(M_{\rm dyn} = (V_{\rm rot}^{2}+3\sigma_v^{2}) R_{\rm HI} / G\) \citep{1996ApJS..105..269H}, where \(G\) is the gravitational constant. Hence, this yields $M_{\rm dyn}(<2.8\text{ kpc}) \approx 1.3 \times 10^8 ~\rm M_\odot$ for J1351+0039. The resulting a dark-to-baryonic mass ratio is $M_{\rm dyn}/M_{\rm bar} \sim 200$. Although beam smearing certainly circularizes the observed projected profile (\(D_{\rm min}/D_{\rm maj}\) = 0.94) and attenuates local velocity gradients, the global linewidth (bottom panel of Figure 1) sets an upper bound on projected rotation of 13.8 \kms. At these low column densities, photoheating by the cosmic UV background is expected to maintain a thermal floor of $T \sim 10^4\text{ K}$ ($\sigma_{\rm th} \approx 10\text{ km s}^{-1}$). We thus conclude that the resulting hydrostatic aspect ratio ($h/R_{\rm HI} \sim \sigma_{\rm th}/V_{\rm circ} \sim 0.6\text{--}0.8$) proves that the system is an intrinsically thick, pressure-supported spheroid rather than an inclined thin disk. The small observed velocity gradient or weak rotation thus contributes negligibly to the overall support, confirming that the system is largely stabilized by dispersion.

For regions without detected optical sources, the sky background dominates the flux measurement \citep{2021AJ....162..274L}. To quantify our sensitivity to low-surface-brightness stellar components, we computed surface brightness limits using the \texttt{sbcontrast} code \citep{2022ApJ...935..160K}. Following standard practice in studies of nearby galaxies, we adopted \(10^{\prime\prime} \times 10^{\prime\prime}\) boxes as representative angular scales for extended sources \citep{2020A&A...644A..42R,2023A&A...671A.141M,2023A&A...669A.103M}. The resulting \(3\sigma\) surface brightness limits are 28.1 and 27.5 mag arcsec\(^{-2}\) in the DESI \(g\) and \(r\) bands, respectively. For irregular galaxies, a tight correlation exists between \HI mass and the optical disk diameter \(D_{25}\), defined at the 25 mag arcsec\(^{-2}\) isophote \citep{1997A&A...324..877B}, expressed as \(\log_{10}(M_{\rm HI}) = 7.0 + 1.95\,\log_{10}(D_{25})\). Applying this relation, the predicted \(D_{25}\) for J1351+0039 is 0.21 kpc. Using this physical scale and \(3\sigma\) surface brightness limits, we derived the limiting apparent magnitude, $m_{g} \approx 20.7$ mag and $m_{r} \approx 20.1$ mag. The corresponding upper limit on the stellar mass ($M_{\star}$) is estimated using the mass-to-light ratio relation $\rm log(\it M_{\star}/L_{g})= \rm -0.601+1.294 \it(m_{g}-m_{r})$, where $L_{g}$ is the $g$-band luminosity \citep{2016AJ....152..177H}. Hence, we derived an upper $g$-band luminosity limit of \(L_g \lesssim 7.5\times10^{4} \, L_\odot\) and an upper stellar mass limit of \(M_\star \lesssim 1.2\times10^{5} \, M_\odot\) for J1351+0039. For comparison, the ultra-faint dwarf Leo T has $M_{\rm HI} \approx 4.1 \times 10^5 M_\odot$ and $M_\star\approx 2.0 \times 10^5 M_\odot$ \citep{2018A&A...612A..26A}. At $3.6\text{ Mpc}$, a diffuse stellar population comparable to Leo T would have a mean surface brightness of $\mu_g \gtrsim 28.5\text{ mag arcsec}^{-2}$, remaining undetectable in DESI data  \citep[see, e.g.,][]{2024ApJ...973...61B}. Deeper space-based imaging (e.g., with HST or JWST) is required to probe for resolved stars and confirm whether the system is truly starless.

J1351+0039 is dominated by dark matter and shows no optical counterpart down to a \(g\)-band surface brightness limit of 28.1 mag arcsec\(^{-2}\), making this isolated system an ideal candidate for either a RELHIC or an ultra-faint galaxy. Its positive systemic velocity, narrow line width ($W_{50}\simeq$27.6 \kms), nearly spherical morphology with an axis ratio of \((D_{\rm min}/D_{\rm maj}\) = 0.94) are consistent with the defining characteristics of a RELHIC\footnote{RELHICs are expected to display aspect ratios > 0.8 and  $W_{50}\simeq$22.0 \kms}. \citep{2017MNRAS.465.3913B}. However, J1351+0039 appear to exhibit a weak rotating gas structure, which, while in principle unexpected for RELHICs, is subdominant to the system.  Since the rotation velocity varies with radius and is very small in the central regions, internal pressure likely provides substantial support for the gas. The pressure-supported nature therefore suggests that J1351+0039 may be a RELHIC that has gradually acquired angular momentum. For isolated systems, thermal pressure  may dominate. Based on the velocity dispersion, we estimate the kinetic temperature as \(T = m_{\rm H}\sigma_v^{2}/k\), where \(m_{\rm H}\) is the hydrogen atomic mass and \(k\) is the Boltzmann constant. The derived temperature of approximately \(1.1\times10^4\) K places the \HI gas in the warm neutral medium regime, in thermal equilibrium with the cosmic ultraviolet background, and has never cooled sufficiently to form stars. While earlier candidates like Cloud-9 (near M94) and those near M51 suffer from environmental and tidal effects due to nearby massive hosts \citep[see e.g.,][]{2024ApJ...973...61B,2026MNRAS.550g1293Z}, the complete isolation of J1351+0039 establishes it as the cleanest and most compelling ``isolated’’ RELHIC candidate identified to date.


\begin{acknowledgements}
   We thank the referees for insightful comments that improved the clarity of this manuscript.  We acknowledge the supports of the National Key R$\&$D Program of China (No. 2025YFE202300) and the National SKA program of China (No. 2025SKA0150101). This work is also supported by the National Natural Science Foundation of China (Grant Nos. 12373001, 12225303, 12421003), the Chinese Academy of Sciences Project for Young Scientists in Basic Research, grant no. YSBR-063. ABL acknowledges support by the Italian Ministry for Universities (MUR) program “Dipartimenti di Eccellenza 2023-2027”
within the Centro Bicocca di Cosmologia Quantitativa (BiCoQ), and support by UNIMIB’s Fondo Di Ateneo Quota Competitiva (project 2024-ATEQC-0050).
\end{acknowledgements}

\bibliographystyle{aa} 
\bibliography{sample631} 
\end{document}